  \documentstyle[twoside,epsf]{article} 
\newcommand{\ep}{\epsilon}
\newcommand{\wt}{\widetilde}
   \catcode`\@=11 
   \long\def\@makefntext#1{ 
   \protect\noindent \hbox to 3.2pt {\hskip-.9pt 
   $^{{\eightrm\@thefnmark}}$\hfil}#1\hfill} 

   \def\@makefnmark{\hbox to 0pt{$^{\@thefnmark}$\hss}} 

   \def\ps@myheadings{\let\@mkboth\@gobbletwo 
   \def\@oddhead{\hbox{} 
   \rightmark\hfil\eightrm\thepage} 
   \def\@oddfoot{}\def\@evenhead{\eightrm\thepage\hfil 
   \leftmark\hbox{}}\def\@evenfoot{} 
   \def\sectionmark##1{}\def\subsectionmark##1{}} 

   \oddsidemargin=\evensidemargin 
   \newcounter{sectionc}\newcounter{subsectionc}\newcounter{subsubsectionc} 
   \renewcommand{\section}[1] {\vspace{12pt}\addtocounter{sectionc}{1} 
   \setcounter{subsectionc}{0}\setcounter{subsubsectionc}{0}\noindent 
   {\tenbf\thesectionc. #1}\par\vspace{5pt}} 
   \renewcommand{\subsection}[1] {\vspace{12pt}\addtocounter{subsectionc}{1} 
   \setcounter{subsubsectionc}{0}\noindent 
   {\bf\thesectionc.\thesubsectionc. {\kern1pt \bfit #1}}\par\vspace{5pt}} 
   \renewcommand{\subsubsection}[1] {\vspace{12pt}\addtocounter{subsubsectionc}{1} 
   \noindent{\tenrm\thesectionc.\thesubsectionc.\thesubsubsectionc. 
   {\kern1pt \tenit #1}}\par\vspace{5pt}} 
   \newcommand{\nonumsection}[1] {\vspace{12pt}\noindent{\tenbf #1} 
   \par\vspace{5pt}} 

   \newcounter{appendixc} 
   \newcounter{subappendixc}[appendixc] 
   \newcounter{subsubappendixc}[subappendixc] 
   \renewcommand{\thesubappendixc}{\Alph{appendixc}.\arabic{subappendixc}} 
   \renewcommand{\thesubsubappendixc} 
   {\Alph{appendixc}.\arabic{subappendixc}.\arabic{subsubappendixc}} 

   \renewcommand{\appendix}[1] {\vspace{12pt} 
   \refstepcounter{appendixc} 
   \setcounter{figure}{0} 
   \setcounter{table}{0} 
   \setcounter{lemma}{0} 
   \setcounter{theorem}{0} 
   \setcounter{corollary}{0} 
   \setcounter{definition}{0} 
   \setcounter{equation}{0} 
   \renewcommand{\thefigure}{\Alph{appendixc}.\arabic{figure}} 
   \renewcommand{\thetable}{\Alph{appendixc}.\arabic{table}} 
   \renewcommand{\theappendixc}{\Alph{appendixc}} 
   \renewcommand{\thelemma}{\Alph{appendixc}.\arabic{lemma}} 
   \renewcommand{\thetheorem}{\Alph{appendixc}.\arabic{theorem}} 
   \renewcommand{\thedefinition}{\Alph{appendixc}.\arabic{definition}} 
   \renewcommand{\thecorollary}{\Alph{appendixc}.\arabic{corollary}} 
   \renewcommand{\theequation}{\Alph{appendixc}.\arabic{equation}} 
   \noindent{\tenbf Appendix \theappendixc #1}\par\vspace{5pt}} 
   \newcommand{\subappendix}[1] {\vspace{12pt} 
   \refstepcounter{subappendixc} 
   \noindent{\bf Appendix \thesubappendixc. {\kern1pt \bfit #1}} 
   \par\vspace{5pt}} 
   \newcommand{\subsubappendix}[1] {\vspace{12pt} 
   \refstepcounter{subsubappendixc} 
   \noindent{\rm Appendix \thesubsubappendixc. {\kern1pt \tenit #1}} 
   \par\vspace{5pt}} 

   \newcommand{\textlineskip}{\baselineskip=13pt} 
   \newcommand{\smalllineskip}{\baselineskip=10pt} 

   \newcommand{\copyrightheading}[1] 
   {\vspace*{-2.5cm}\smalllineskip{\flushleft 
   }} 

   \def\abstracts#1#2#3{{ 
   \centering{\begin{minipage}{4.5in}\footnotesize\baselineskip=10pt 
   \parindent=0pt #1\par 
   \parindent=15pt #2\par 
   \parindent=15pt #3 
   \end{minipage}}\par}} 

   \def\keywords#1{{ 
   \centering{\begin{minipage}{4.5in}\footnotesize\baselineskip=10pt 
   {\footnotesize\it Keywords}\/: #1 
   \end{minipage}}\par}}

   \renewenvironment{thebibliography}[1] 
   {\frenchspacing 
   \ninerm\baselineskip=11pt 
   \begin{list}{\arabic{enumi}.} 
   {\usecounter{enumi}\setlength{\parsep}{0pt} 
   \setlength{\leftmargin 12.7pt}{\rightmargin 0pt}
   \setlength{\itemsep}{0pt} \settowidth 
   {\labelwidth}{#1.}\sloppy}}{\end{list}} 

   \newcounter{itemlistc} 
   \newcounter{romanlistc} 
   \newcounter{alphlistc} 
   \newcounter{arabiclistc}

   \newcommand{\fcaption}[1]{ 
   \refstepcounter{figure} 
   \setbox\@tempboxa = \hbox{\footnotesize Fig.~\thefigure. #1} 
   \ifdim \wd\@tempboxa > 5in 
   {\begin{center} 
   \parbox{5in}{\footnotesize\smalllineskip Fig.~\thefigure. #1} 
   \end{center}} 
   \else 
   {\begin{center} 
   {\footnotesize Fig.~\thefigure. #1} 
   \end{center}} 
   \fi} 

   \newcommand{\tcaption}[1]{ 
   \refstepcounter{table} 
   \setbox\@tempboxa = \hbox{\footnotesize Table~\thetable. #1} 
   \ifdim \wd\@tempboxa > 5in 
   {\begin{center} 
   \parbox{5in}{\footnotesize\smalllineskip Table~\thetable. #1} 
   \end{center}} 
   \else 
   {\begin{center} 
   {\footnotesize Table~\thetable. #1} 
   \end{center}} 
   \fi} 

   \def\pmb#1{\setbox0=\hbox{#1} 
   \kern-.025em\copy0\kern-\wd0 
   \kern.05em\copy0\kern-\wd0 
   \kern-.025em\raise.0433em\box0}

   \def\fnt#1#2{\footnotetext{\kern-.3em 
   {$^{\mbox{\scriptsize #1}}$}{#2}}} 

   \def\fpage#1{\begingroup 
   \voffset=.3in 
   \thispagestyle{empty}\begin{table}[b]\centerline{\footnotesize #1} 
   \end{table}\endgroup} 

   \def\runninghead#1#2{\pagestyle{myheadings} 
   \markboth{{\protect\footnotesize\it{\quad #1}}\hfill} 
   {\hfill{\protect\footnotesize\it{#2\quad}}}} 
   \font\tenrm=cmr10 
   \font\tenit=cmti10 
   \font\tenbf=cmbx10 
   \font\bfit=cmbxti10 at 10pt 
   \font\ninerm=cmr9

   \font\eightrm=cmr8

   \def\FigName{figure}%
   \newbox\captionbox 
   \long\def\@makecaption#1#2{%
   \ifx\FigName\@captype 
   \vskip\abovecaptionskip 
   \setbox\tempbox\hbox{{\figurecaptionfont #1\hskip1em #2}} 
   \ifdim\wd\tempbox< 28pc 
   \centerline{\box\tempbox} 
   \else 
   {\figurecaptionfont #1\hskip1em #2\par} 
   \fi\else 
   \setbox\tempbox\hbox{{\tablecaptionfont #1\hskip1em #2}} 
   \ifdim\wd\tempbox< 28pc 
   \centerline{\box\tempbox} 
   \else 
   {\tablecaptionfont #1\hskip1em #2\par}%
   \fi 
   \vskip\belowcaptionskip 
   \fi} 
   \InputIfFileExists{psfig.sty} 
   {\typeout{^^Jpsfig.sty inputed...ok}}{\typeout{^^JWarning: psfig.sty could be be found.^^J}} 
   \InputIfFileExists{epsfsafe.tex} 
   {\typeout{^^Jepsfsafe.tex inputed...ok}} 
   {\typeout{^^JWarning: epsfsafe.tex could not be found.^^J}} 
   \InputIfFileExists{epsfig.sty} 
   {\typeout{^^Jepsfig.sty inputed...ok}}{\typeout{^^JWarning: epsfig.sty could not be found.^^J}} 
   \InputIfFileExists{epsf.sty} 
   {\typeout{^^Jepsf.sty inputed...ok}}{\typeout{^^JWarning: epsf.sty could not be found.^^J}}%
   \def\fps@figure{tbp} 
   \def\ftype@figure{1} 
   \def\ext@figure{lof} 
   \def\fnum@figure{Fig.\ \thefigure} 
   \def\qed{\hbox{${\vcenter{\vbox{ 
   \hrule height 0.4pt\hbox{\vrule width 0.4pt height 6pt 
   \kern5pt\vrule width 0.4pt}\hrule height 0.4pt}}}$}} 

\begin{document} 
   \setlength{\textheight}{8.0truein} 

   \runninghead{Purification of entangled coherent states} 
   {H. Jeong and M. S. Kim} 

   \normalsize\textlineskip 
   \thispagestyle{empty} 
   \setcounter{page}{1} 

   \copyrightheading{} 

   \vspace*{0.88truein} 

   \fpage{1} 
   \centerline{\bf 
 PURIFICATION OF ENTANGLED COHERENT STATES} 
   \vspace*{0.37truein} 
   \centerline{\footnotesize 
   H. JEONG and M. S. KIM} 
   \vspace*{0.015truein} 
   \centerline{\footnotesize\it
School of Mathematics and Physics, Queen's University} 
   \baselineskip=10pt 
   \centerline{\footnotesize\it  Belfast BT7 1NN, United Kingdom} 
   \vspace*{0.225truein} 

   \vspace*{0.21truein} \abstracts{ We suggest an entanglement
     purification scheme for mixed entangled coherent states using
     50-50 beam splitters and photodetectors.  This scheme is directly
     applicable for mixed entangled coherent states of the Werner
     type, and can be useful for general mixed states using additional
     nonlinear interactions.  We apply our scheme to
     entangled coherent states decohered in a vacuum environment and find the
     decay time until which they can be purified.}{}{} \vspace*{10pt}
    \vspace*{3pt}

   \vspace*{1pt}\textlineskip 
   \section{Introduction} 
   \vspace*{-0.5pt} 
   \noindent

Entanglement is an important manifestation of quantum mechanics.
Highly entangled states play a key role in an efficient realization of
quantum information processing including quantum teleportation
\cite{Bennett93}, cryptography \cite{Ekert91} and computation
\cite{Barenco95}.  When an entangled state prepared for quantum
information processing is open to an environment, the pure entangled
state becomes mixed one and the entanglement of the original state
becomes inevitably degraded.  To obtain highly entangled
states from less entangled mixed ones, entanglement purification
protocols \cite{Bennett96,BBPS96,Pan} have been proposed.

Recently, entangled coherent states \cite{Sanders} have been studied
for quantum information processing and 
nonlocality test \cite{Enk,JKL01,Wang,JK,MMS,Rice,Derek}.  Teleportation schemes
via entangled coherent states \cite{Enk,JKL01,Wang} and
quantum computation with coherent-state qubits \cite{JK,Ralph} using multi-mode
entangled coherent states \cite{JK} have been investigated.  
These suggestions \cite{Enk,JKL01,Wang,JK} require highly
entangled coherent states for successful realization.  Even though
entanglement concentration for partially entangled pure states has
been studied \cite{JKL01}, there is a need for a purification scheme for
mixed states.

In this paper, we suggest an entanglement purification scheme for
mixed entangled coherent states.  This scheme is based on the use of
50-50 beam splitters and photodetectors.  We show that our scheme can
be directly applied for entangled coherent states of the Werner form
based on quasi-Bell states \cite{Hirota01!}.  The scheme can also be
useful for general mixed entangled coherent states using additional
nonlinear interactions.

In Sec.~2, we review entangled coherent states and their
characteristics with a entanglement concentration scheme for pure
states.
The purification scheme for mixed states is suggested and applied
to a simple
example in Sec.~3.
We also discuss how this scheme is related to previously suggested
ones \cite{Bennett96,Pan}, and
we show that our scheme is applicable to any Werner-type states.  The
required bilateral operations for general mixed states are discussed
in Sec.~4.  In Sec.~5, we apply our scheme to decohered entangled
coherent states in a vacuum environment.  Finally, we present an
example of application for multi-mode entanglement purification.

   \section{Entangled coherent states and pure state concentration} 
 \label{sec2}  \noindent 
We define entangled coherent states
\begin{eqnarray}
\label{eq:ecs1}
&&|\Phi_\varphi\rangle_{ab}=N_\varphi(|\alpha\rangle_a|\alpha\rangle_b+e^{i\varphi}
 |-\alpha\rangle_a|-\alpha\rangle_b),\\
\label{eq:ecs1p}
&&|\Psi_\varphi\rangle_{ab}=N_\varphi(|\alpha\rangle_a|-\alpha\rangle_b+e^{i\varphi} |-\alpha\rangle_a|\alpha\rangle_b),
\end{eqnarray}
where $\alpha=\alpha_r+i\alpha_i$ is the complex amplitude of the
coherent state $|\alpha\rangle$, $\varphi$ is a real local phase
factor, and $N_\varphi=\{2(1+\cos\varphi
e^{-4|\alpha|^2})\}^{-1/2}$ is the normalization factor.  Entangled coherent states
(\ref{eq:ecs1}) and (\ref{eq:ecs1p}) can be generated using a
nonlinear medium and lossless 50-50 beam splitter \cite{Yurke}.  A
coherent superposition state (cat state) can be generated from a
coherent state $|\pm\sqrt{2}\alpha\rangle$ by a nonlinear interaction
in a Kerr medium \cite{Yurke}.  When the coherent superposition state
$M_\varphi(|\sqrt{2}\alpha\rangle+e^{i\varphi}|-\sqrt{2}\alpha\rangle)$,
where $M_\varphi$ is the normalization factor, is input to a 50-50
beam splitter while nothing is input to the other input port, the resulting state is an entangled coherent state
(\ref{eq:ecs1}) or (\ref{eq:ecs1p}) depending on the relative phase
between the reflected and transmitted fields from the beam splitter.

However, Kerr nonlinearity of currently available nonlinear media is
too small to generate the required coherent superposition state.
 It was pointed
out that one needs optical fiber of about 3,000km to generate a
coherent superposition state with currently available Kerr
nonlinearity \cite{Gerry}.  Even though it is possible to make such a
long nonlinear optical fiber, the decoherence effect
during the propagation is too large.  Some alternative methods have
been studied to generate a superposition of macroscopically
distinguishable states using conditional measurements
\cite{Song,Dakna}.  One drawback of these schemes is
due to the low efficiency of photon-number measurement.
 Cavity
quantum electrodynamics has been studied to enhance nonlinearity
\cite{Tu}, and there have been experimental demonstrations of generating cat
stated in a cavity and in a trap \cite{MB,Mon}.  Unfortunately, all the
suggested schemes for quantum information processing with coherent
states \cite{Enk,JKL01,Wang,JK,Ralph}, including the work in this
paper, require {\it propagating} optical cat states.
There were other suggestions to generate cat states
with trapped ions \cite{Solano} and with solitons \cite{Nat}.

Electromagnetically induced transparency (EIT) has been studied as
a method to obtain giant Kerr nonlinearity \cite{Lukin}.
There was an experimental report of an indirect measurement of a giant
Kerr nonlinearity utilizing EIT \cite{giant}, but
this developing technology has not been exactly at hand to generate an
optical cat state.
In short, generating a propagating optical cat state is extremely
demanding and difficult with currently available Kerr effect, so that
a scheme to generate a cat state with small nonlinear effect needs
to be studied.  In this paper, we assume ideal nonlinear effect to
generate a coherent superposition state while its generating scheme
with small nonlinearity is being studied and will be demonstrated
elsewhere \cite{comming}.

It is possible to define a 2-dimensional Hilbert space $\cal H_\alpha$
with two linear independent vectors $|\alpha\rangle$ and
$|-\alpha\rangle$.  For example, an orthonormal basis for Hilbert
space $\cal H_\alpha$ can be constructed 
\begin{eqnarray}
\label{C11}
&&|u\rangle=M_+(|\alpha\rangle+|-\alpha\rangle),\\
\label{C22}
&&|v\rangle=M_-(|\alpha\rangle-|-\alpha\rangle),
\end{eqnarray}
where $M_+$ and $M_-$ are normalization factors.  Using the orthogonal
basis, we can study the entangled coherent state in the
$2\times2$-dimensional Hilbert space. For $\varphi=0$,
$|\Phi_\varphi\rangle$ can be represented as
\begin{equation}
|\Phi_{\varphi=0}\rangle=\frac{N(\varphi=0)}{2M_+^2}\Big(|u\rangle|u\rangle+\frac{M_+^2}{M_-^2}|v\rangle|v\rangle\Big).
\end{equation}

The entanglement of $|\Phi_\varphi\rangle$ and $|\Psi_\varphi\rangle$
can be quantified by the von Neumann entropies of their reduced
density matrices.
We find that the degree of entanglement  $E(|\alpha|,\varphi)$ for
$|\Phi_\varphi\rangle$ and $|\Psi_\varphi\rangle$ are the same and
\begin{eqnarray}
  &&E(|\alpha|,\varphi)=-\frac{N_\varphi^2}{\ln2}\Big\{{\cal
    M}(0){\cal M}(\varphi)\ln[N_\varphi^2 {\cal M}(0){\cal M}(\varphi)] \nonumber \\ 
  &&~~~~~~~~~~~~~~~~~~~~~~~~~+{\cal M}(\pi){\cal M}(\varphi+\pi)\ln[N_\varphi^2{\cal
    M}(\pi){\cal M}(\varphi+\pi)]\Big\},
\end{eqnarray}
where ${\cal M}(\varphi)=1+\cos\varphi e^{-2|\alpha|^2}$.  Note that
$E(|\alpha|,\varphi)$ is the degree of entanglement defined not in
continuous variables as in \cite{Parker} but in the $2\times2$ space ${\cal
  H}_\alpha^{(1)}\otimes{\cal H}_\alpha^{(2)}$.  The degree of
entanglement $E(|\alpha|,\varphi)$ varies not only by the coherent
amplitude $\alpha$ but also by the relative phase $\varphi$.  When
$\varphi=\pi$, both the entangled coherent states
$|\Phi_\varphi\rangle$ and $|\Psi_\varphi\rangle$ are maximally
entangled regardless of $\alpha$, {\it i.e.}, $E(|\alpha|,\pi)=1$.  When
$\varphi=0$, on the other hand, $E(|\alpha|,\varphi)$ is minimized for a
given coherent amplitude $\alpha$.
These characteristics of entangled coherent states have already been
pointed out by some authors \cite{Hirota01!,XWang}.

\begin{figure} [htbp] 
  \vspace*{13pt} \centerline{\psfig{file=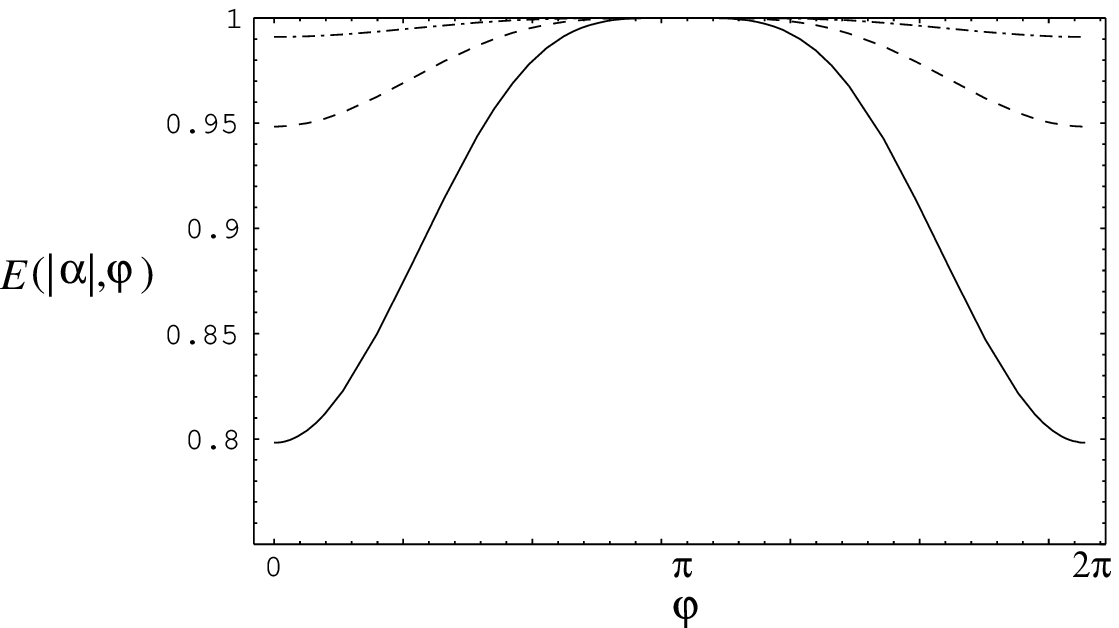,
      width=6.2cm}} \vspace*{13pt} \fcaption{ Measure of entanglement
    $E(|\alpha|,\varphi)$, quantified by the von Neumann entropy of
    the reduced density matrix, against the relative phase $\varphi$
    of the entangled coherent state.  $|\alpha|=0.8$ (solid line),
    $|\alpha|=1$ (dashed line), $|\alpha|=1.2$ (dot-dashed line), and
    $0\leq \varphi< 2\pi$.  This figure shows that when $|\alpha|$ is
    large, the quasi-Bell states are good approximations to maximally
    entangled Bell states.
\label{fig:moe} } 
   \end{figure}

Substituting $\varphi$ by 0 and $\pi$, we define  quasi-Bell states \cite{Hirota01!}
\begin{eqnarray}
\label{eq:qbs1}
&&|\Phi_\pm\rangle_{ab}=N_\pm(|\alpha\rangle_a|\alpha\rangle_b\pm|-\alpha\rangle_a|-\alpha\rangle_b),\\
\label{eq:qbs2}
&&|\Psi_\pm\rangle_{ab}=N_\pm(|\alpha\rangle_a|-\alpha\rangle_b\pm|-\alpha\rangle_a|\alpha\rangle_b).
\end{eqnarray}
These states are orthogonal to each other except
\begin{equation}
\langle\Psi_+|\Phi_+\rangle=\frac{1}{\cosh 2|\alpha|^2}.
\end{equation}
We immediately see that as $|\alpha|$ grows, they rapidly become
orthogonal. 
In Fig.~\ref{fig:moe}, we also show that the entanglement
$E(|\alpha|,\varphi)$ drastically approaches to 1 as $|\alpha|$ increases.
We calculate  $E(2,0)\simeq0.9999997$ and
$E(3,0)\simeq1-6.7\times10^{-16}$, which shows quasi-Bell states are
good approximations to maximally entangled Bell states.

It was shown that complete Bell-state measurements on a product
Hilbert space of two two-level systems are not possible using linear
elements \cite{L}. However, a remarkable feature of quasi-Bell
states is that each one of them can be unambiguously discriminated using only
linear elements.  Suppose that each mode of the entangled state is
incident on a 50-50 beam splitter. After passing the beam splitter, the
quasi-Bell states become
\begin{eqnarray}
\label{fig:setup1}
|\Phi_+\rangle_{ab} &\longrightarrow& 
  |U\rangle_f|0\rangle_g,
  \nonumber \\
  |\Phi_-\rangle_{ab} &\longrightarrow &
  |V\rangle_f|0\rangle_g,
  \nonumber \\
  |\Psi_+\rangle_{ab} &\longrightarrow &
  |0\rangle_f|U\rangle_g, 
  \nonumber \\
  |\Psi_-\rangle_{ab} &\longrightarrow &
  |0\rangle_f|V\rangle_g,
\end{eqnarray}
where the even cat state $|U\rangle=M^\prime_+(|\sqrt{2}\alpha\rangle
+|-\sqrt{2}\alpha\rangle)$ with the normalization factor $M^\prime_+$
contains only even numbers of photons, while the odd cat state
$|V\rangle=M^\prime_-(|\sqrt{2}\alpha\rangle
-|-\sqrt{2}\alpha\rangle)$ with the normalization factor $M^\prime_-$
contains only odd numbers of photons \cite{JKL01}.  By setting two
photodetectors for the output modes $f$ and $g$ respectively to perform number
parity measurement, 
the quasi-Bell measurement can be simply achieved.  For example, if
an odd number of photons is detected for mode $f$, the state
$|\Phi_-\rangle$ is measured, and if an odd number of photons is
detected for mode $g$, then $|\Psi_-\rangle$ is measured.  Even though
there is non-zero probability of failure in which both of the
detectors do not register a photon due to the non-zero overlap of
$|\langle0|U\rangle|^2=e^{-2|\alpha|^2}/(1+e^{-4|\alpha|^2})^2$, the
failure probability is very small for an appropriate choice of
$\alpha$ and the failure is known from the result whenever it occurs.
This quasi-Bell measurement scheme can be used for concentration of
pure entangled coherent states \cite{JKL01}.

   \section{Entanglement purification for mixed states}
   \noindent 

\begin{figure} [htbp] 
   \vspace*{13pt} \centerline{\psfig{file=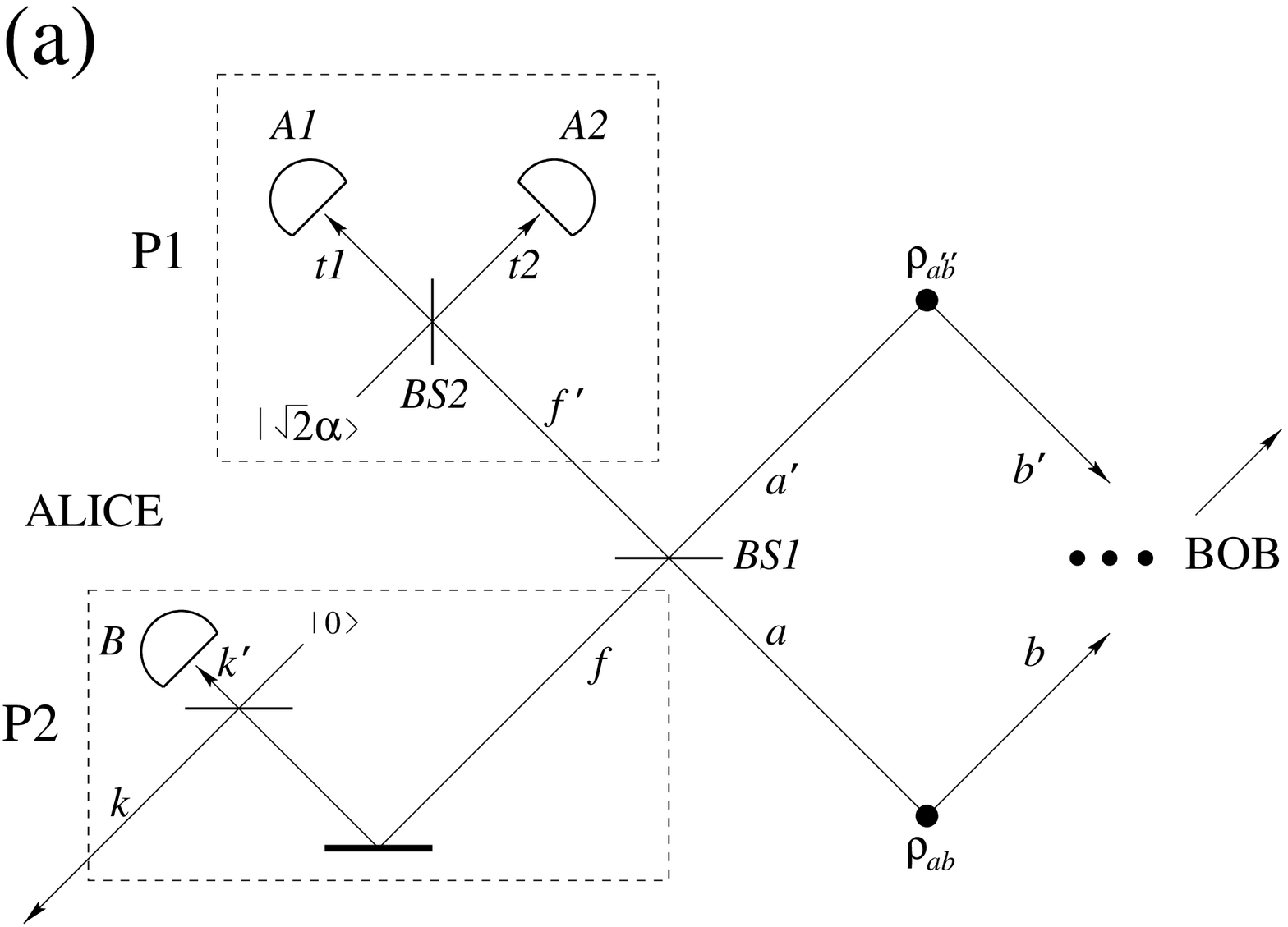, width=8.2cm}}
   \vspace*{13pt} \centerline{\psfig{file=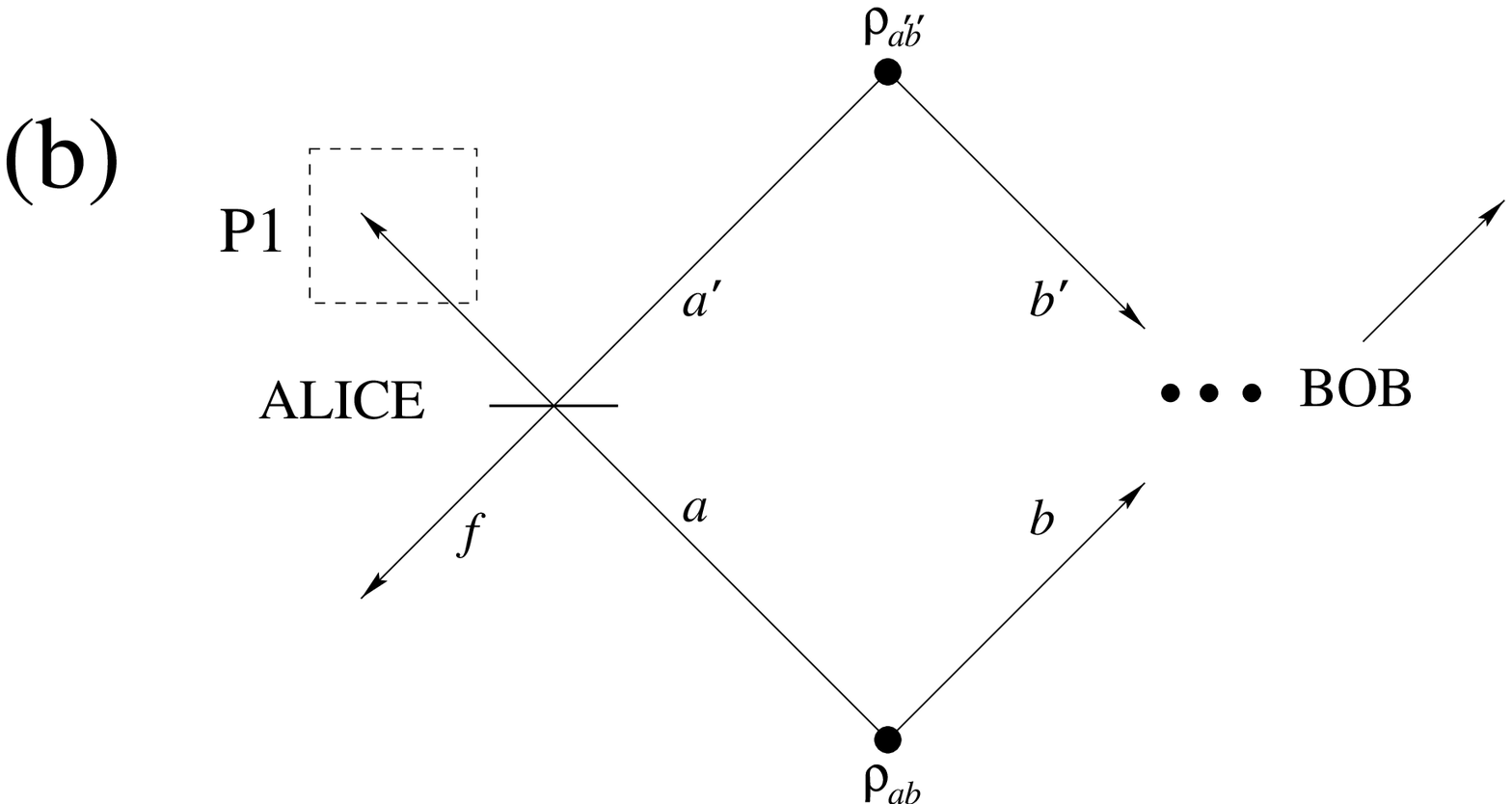, width=8.2cm}}
   \vspace*{13pt} \fcaption{ (a) Entanglement purification scheme for
     mixed entangled coherent states.  P1 tests if the incident fields
     $a$ and $a^\prime$ were in the same state by simultaneous clicks
     at $A1$ and $A2$.  For P2, detector $B$ is set for photon parity
     measurement.  Bob performs the same on his field of modes $b$ and
     $b^\prime$ as Alice.  If Alice and Bob measure the same parity,
     the pair is selected.  By iterating this process maximally
     entangled pairs can be obtained from a sufficiently large
     ensemble of mixed states.  (b) Simpler purification scheme to
     increase the coherent amplitude of the purified state.  The
     success probability of this scheme is more than twice as large as the
     scheme with P1 and P2 shown in (a).
\label{fig:ep} } 
   \end{figure}

Suppose that Alice and Bob's ensemble to be purified is
represented by
\begin{equation}
\label{ensemble}
\rho_{ab}=F|\Phi_-\rangle\langle\Phi_-|+(1-F)|\Psi_-\rangle\langle\Psi_-|,
\end{equation}
where $F$ is the fidelity defined as $\langle\Phi_-|\rho_{ab}|\Phi_-\rangle$ and $0<F<1$.
Note that $|\Phi_-\rangle$ and $|\Psi_-\rangle$ are maximally
entangled and orthogonal to each other regardless of $\alpha$.  Alice and
Bob choose two pairs from the ensemble which are
represented by the following density operator 
\begin{eqnarray}
\label{eq:rho*2}
&&\rho_{a b} \rho_{a^\prime b^\prime}=F^2|\Phi_-\rangle\langle\Phi_-|\otimes|\Phi_-\rangle\langle\Phi_-|
+F(1-F)|\Phi_-\rangle\langle\Phi_-|\otimes|\Psi_-\rangle\langle\Psi_-|\nonumber\\
&&~~~~~~~~~~~~+F(1-F)|\Psi_-\rangle\langle\Psi_-|\otimes|\Phi_-\rangle\langle\Phi_-|+(1-F)^2|\Psi_-\rangle\langle\Psi_-|\otimes|\Psi_-\rangle\langle\Psi_-|.
\end{eqnarray}
The fields of modes $a$ and $a^\prime$ are in Alice's possession while
$b$ and $b^\prime$ in Bob's. In Fig.~\ref{fig:ep}(a), we show that
Alice's action to purify the mixed entangled state. The same is
conducted by Bob on his fields of $b$ and $b^\prime$.

There  are four possibilities for the fields of $a$ and $a^\prime$
incident onto the beam splitter ($BS1$), which gives the output
(In the following, only the cat part for a component of the mixed
state is shown to describe the action of the apparatuses)
\begin{eqnarray}
\label{eq:tr1}
&&|\alpha\rangle_a|\alpha\rangle_{a^\prime}\longrightarrow|\sqrt{2}\alpha\rangle_f|0\rangle_{f^\prime},\\
&&|\alpha\rangle_a|-\alpha\rangle_{a^\prime}\longrightarrow|0\rangle_f|\sqrt{2}\alpha\rangle_{f^\prime},\label{eq:tr2}\\
&&|-\alpha\rangle_a|\alpha\rangle_{a^\prime}\longrightarrow|0\rangle_f|-\sqrt{2}\alpha\rangle_{f^\prime},\label{eq:tr3}\\
&&|-\alpha\rangle_a|-\alpha\rangle_{a^\prime}\longrightarrow|-\sqrt{2}\alpha\rangle_f|0\rangle_{f^\prime}\label{eq:tr4}.
\end{eqnarray}
In the boxed apparatus P1, Alice checks if modes $a$ and $a^\prime$
were in the same state by counting photons at the photodetectors $A1$
and $A2$. If both modes $a$ and $a^\prime$ are in $|\alpha\rangle$
or $|-\alpha\rangle$, $f^\prime$ is in the vacuum, in which case the
output field of the beam splitter $BS2$ is
$|\alpha,-\alpha\rangle_{t1, t2}$.  Otherwise, the output field is
either $|2\alpha,0\rangle_{t1, t2}$ or $|0,2\alpha\rangle_{t1, t2}$.
When both the photodetectors $A1$ and $A2$ register any photon(s),
Alice and Bob are sure that the two modes $a$ and $a^\prime$ were in
the same state but when either $A1$ or $A2$ does not resister a
photon, $a$ and $a^\prime$ were likely in different states. Of course,
there is a probability not to resister a photon even though the two
modes were in the same state, which is due to the nonzero overlap of
$|\langle0|\sqrt{2}\alpha\rangle|^2$.

It can be simply shown that the second and third terms of
Eq.~(\ref{eq:rho*2}) are always discarded by the action of P1 and
Bob's apparatus same as P1.
For example, at the output ports of $BS1$ and Bob's beam splitter
corresponding to $BS1$, $|\Phi_-\rangle_{ab}|\Psi_-\rangle_{a^\prime
  b^\prime}$ becomes
\begin{eqnarray}
&&|\Phi_-\rangle_{ab}|\Psi_-\rangle_{a^\prime b^\prime}  
 \longrightarrow N_-^2\big(|\sqrt{2}\alpha,0,0,\sqrt{2}\alpha\rangle-|0,\sqrt{2}\alpha,\sqrt{2}\alpha,0\rangle
  \nonumber\\ 
&&~~~~~~~~~~~~~~~~~~~~~~~~~~~~~~~-|0,-\sqrt{2}\alpha,-\sqrt{2}\alpha,0\rangle+|-\sqrt{2}\alpha,0,0,-\sqrt{2}\alpha\rangle\big)_{fgf^\prime
  g^\prime},
\label{bs2}
\end{eqnarray}
where $g$ and $g^\prime$ are the output field modes from Bob's beam
splitter corresponding to $BS1$.
The fields of modes $f^\prime$ and $g^\prime$ can never be in
$|0\rangle$ at the same time; at least, one of the four detectors of
Alice and Bob must not click.  The third term of Eq.~(\ref{eq:rho*2}) can be shown to lead
to the same result by the same analysis.

For the cases of the first and fourth terms in Eq.~(\ref{eq:rho*2}), all four detectors
may register photon(s).
After the beam splitter, the ket of
$(|\Phi_-\rangle\langle\Phi_-|)_{ab}\otimes
(|\Phi_-\rangle\langle\Phi_-|)_{a^\prime b^\prime}$ 
 of Eq.~(\ref{eq:rho*2}) becomes
\begin{equation}
\label{+++}
|\Phi_-\rangle_{ab}|\Phi_-\rangle_{a^\prime b^\prime} 
\longrightarrow
|\Phi^\prime_+\rangle_{fg}|0,0\rangle_{f^\prime g^\prime}-|0,0\rangle_{fg}|\Phi^\prime_+\rangle_{f^\prime g^\prime},
\end{equation}
where $|\Phi^\prime_+\rangle
=N_+^\prime(|\sqrt{2}\alpha,\sqrt{2}\alpha\rangle+|-\sqrt{2}\alpha,-\sqrt{2}\alpha\rangle)$
with the normalization factor $N_+^\prime$.  The normalization factor
in the right hand side of Eq.~(\ref{+++}) is omitted.  The first term
is reduced to $|\Phi^\prime_+\rangle_{fg}\langle\Phi^\prime_+|$ after
$|0,0\rangle_{f^\prime g^\prime}\langle0,0|$ is measured out by Alice
and Bob's P1's.  Similarly, the fourth term of Eq.~(\ref{eq:rho*2})
yields $|\Psi^\prime_+\rangle_{fg}\langle\Psi^\prime_+|$, where
$|\Psi^\prime_+\rangle$ is defined in the same way as
$|\Phi_+^\prime\rangle$, after $|0,0\rangle_{f^\prime
  g^\prime}\langle0,0|$ is measured.  Thus the density matrix for the
field of modes $f$ and $g$ conditioned on simultaneous measurement of
photons at all four photodetectors is
\begin{equation}
\label{mid}
\rho_{fg}=F^\prime|\Phi_+^\prime\rangle\langle\Phi_+^\prime|+(1-F^\prime)|\Psi_+^\prime\rangle\langle\Psi_+^\prime|,
\end{equation}
where 
\begin{equation}
\label{1/2}
F^\prime=\frac{F^2}{F^2+(1-F)^2}
\end{equation}
and $F^\prime$ is always larger than $F$ for
any $F>1/2$.

If the pair is selected by Alice and Bob's P1's, each of them performs
another process (P2) for the selected pair.  The pair is incident onto
a 50-50 beam splitter at each site of Alice and Bob shown in
Fig.~\ref{fig:ep}(a). If the selected pair is
$|\Phi_+^\prime\rangle\langle\Phi_+^\prime|$ of Eq.~(\ref{mid}),
then the beam splitter gives
\begin{equation}
|\Phi^\prime_+\rangle_{fg}
\longrightarrow|\Phi_+\rangle_{kl}\Big(\frac{M_-}{M_+}|u,u\rangle_{k^\prime
  l^\prime}+\frac{M_+}{M_-}|v,v\rangle_{k^\prime l^\prime}\Big)
+|\Phi_-\rangle_{kl}\frac{N_+}{N_-}\Big(|u,v\rangle_{k^\prime
  l^\prime}+|v,u\rangle_{k^\prime l^\prime}\Big),
\label{nn}
\end{equation}
where $l$ and $l^\prime$ are field modes at Bob's site corresponding
to $k$ and $k^\prime$.  The normalization factor is
omitted in Eq.~(\ref{nn}).  It is known that $|u\rangle$ contains only
even numbers of photons and $|v\rangle$ contains only odd numbers of
photons.  The state is reduced to $|\Phi_-\rangle$ when
different parities are measured at $k^\prime$ and $l^\prime$ by Alice
and Bob respectively.  The same analysis shows that $|\Psi_-\rangle$
remains by P2's for $|\Psi_+^\prime\rangle_{fg}\langle\Psi_+^\prime|$ of
Eq.~(\ref{mid}) which is originated from the fourth term of
Eq.~(\ref{eq:rho*2}).

  The total state after the full process
becomes
\begin{equation}
\label{eq:rho2}
\rho_{fg}=F^\prime|\Phi_-\rangle\langle\Phi_-|+(1-F^\prime)|\Psi_-\rangle\langle\Psi_-|.
\end{equation}
We already saw from Eq.~(\ref{1/2}) that $F^\prime$ is larger than
$F$ for any $F>1/2$.  Alice and Bob can perform as many iterations as
they need for higher entanglement.  The success probability $P_s$ for
one iteration is
\begin{equation}
P_s=\frac{F^2+(1-F)^2}{4}\Big(1-\frac{2e^{-4|\alpha|^2}}{1+e^{-8|\alpha|^2}}\Big)
\Big(\frac{1-e^{-4|\alpha|^2}}{1+e^{-8|\alpha|^2}}\Big),
\end{equation}
which approaches to
$P_s=\frac{F^2+(1-F)^2}{4}$ and
 $1/8\leq P_s\leq 1/4$ for $|\alpha|\gg1$.
 
By reiterating this process including P1 and P2, Alice and Bob can
distill some maximally entangled states $|\Phi_-\rangle$
asymptotically.  Of course, a sufficiently large ensemble and initial fidelity $F>1/2$
are required for successful purification \cite{Bennett96}. P2 may
be different depending on the type of entangled coherent states to be
distilled.  For example, if Alice and Bob need to distill
$|\Phi_+\rangle$ instead of $|\Phi_-\rangle$, pairs should be
selected when the measurement outcomes yield the same parity.

Let us now consider the roles of P1 and P2 by comparing our scheme
with refs.~\cite{Bennett96} and \cite{Pan}.  Pan {\it et al.}
suggested a purification scheme for the entanglement of linearly
polarized photons, where they use polarizing beam splitters (PBS's)
with photodetectors to test if the two photons are in the same
polarization \cite{Pan}.  From Eqs.~(\ref{eq:tr1}) to (\ref{eq:tr4}),
we pointed out that P1 is to test whether the two fields $a$ and
$a^\prime$ are in the same state.  Hence P1 plays a similar role in
our scheme as PBS's in \cite{Pan}.  Next, consider P2 which enables to
perform orthogonal measurement based on
$|\alpha\rangle\pm|-\alpha\rangle$.  This measurement is also
necessary in the other schemes \cite{Bennett96,Pan}. (We will show
later that this process (P2) is not always necessary in our scheme.)
Pan {\it et al.} explained that a PBS in their scheme has the same
effect as a controlled-NOT gate in the scheme suggested by Bennett
{\it et al.} \cite{Bennett96} except that the success probability is
half as large as \cite{Bennett96}.  Both the schemes
\cite{Bennett96,Pan} can be directly applied to any Werner states
without additional bilateral rotations, thereby it is clear that
our scheme is also applicable to any Werner-type states.

  If Alice and Bob want to distill entangled
coherent states $|\Phi_+\rangle$ or $|\Psi_+\rangle$ while increasing
their coherent amplitudes, it can be simply accomplished by performing
only P1 in Fig.~\ref{fig:ep}(b).  Suppose that Alice and Bob
need to purify a type of ensemble
\begin{equation}
\rho_{ab}=G_1|\Phi_+\rangle\langle\Phi_+|+G_2|\Psi_+\rangle\langle\Psi_+|,
\end{equation}
where $G_1+G_2\simeq1$ for $|\alpha|\gg1$.
If P1 is successful, the selected pair becomes
\begin{equation}
\rho_{fg}=G_1^\prime|\Phi_+^\prime\rangle\langle\Phi_+^\prime|+G_2^\prime|\Psi_+^\prime\rangle\langle\Psi_+^\prime|,
\end{equation}
where $G_1^\prime$ is larger than $G_1$ for any $G_1>G_2$. 
After $n$ iterations, they get a subensemble with higher fidelity of 
\begin{eqnarray}
\label{PhiF}
|\Phi^F_+\rangle&=&{\cal N}_+(|2^{n/2}\alpha\rangle|2^{n/2}\alpha\rangle+|-2^{n/2}\alpha\rangle|-2^{n/2}\alpha\rangle),
\end{eqnarray}
where the coherent amplitude has increased. Here, ${\cal N}_+$ is a
normalization factor.  For example, if
$G_1$ is 2/3 and coherent amplitude $\alpha$ is 2, the fidelity and
the amplitude will be $\sim0.99999$ and $8$ respectively after three
times of iterations.

Note that the sucess probability $P^\prime_s$ of this simplified scheme is
\begin{equation}
P^\prime_s=\frac{F^2+(1-F)^2}{2}\Big(1-\frac{2e^{-4|\alpha|^2}}{1+e^{-8|\alpha|^2}}\Big),
\end{equation}
which is more than twice
as large as that of the scheme shown in Fig.~\ref{fig:ep}(a) and  approaches to  $1/4<P^\prime_s<1/2$ for $|\alpha|\gg1$. 
This is due to the fact that the process P2 is not directly for
entanglement purification differently from the other two schemes
\cite{Bennett96,Pan}.  We separated P1 and P2 while the other schemes
do both processes by one measurement.  In our case, the process P1
purifies the mixed ensemble but the resulting state has a larger
amplitude.
It should be noted that even though the simplified scheme is applicable to any
Werner-type states, (symmetric) entangled coherent states
$|\Phi_+\rangle$ and $|\Psi_+\rangle$ can only be obtained by it.

   \section{Purification for general mixed states} 
   \noindent 

We have shown that a mixed Werner state may be purified
using beam splitters and photodetectors. A general mixed state may be
transformed into a Werner state by random bilateral rotations
\cite{BBPS96,Werner}. The Werner state can then be distilled purified.
For the case of spin-1/2 systems, the
required rotations are $B_x$, $B_y$ and $B_z$ which correspond to
$\pi/2$ rotations around $x$, $y$ and $z$ axes.

The $B_x$ rotation can be realized using a nonlinear medium for the
entangled coherent state.  The
anharmonic-oscillator Hamiltonian of a Kerr medium is
\cite{Yurke}
\begin{equation}
\label{eq:nonlinear}
{\cal H}_{NL}=\hbar\omega a^\dag a+\hbar\Omega(a^\dag a)^2,
\end{equation}
where $\omega$ is the energy level splitting for the
harmonic-oscillator part of the Hamiltonian and $\Omega$ is the
strength of the anharmonic term.  When the interaction time $t$ in the
medium is $\pi/\Omega$, coherent states $|\alpha\rangle$ and
$|-\alpha\rangle$ evolve as follows:
\begin{eqnarray}
&&|\alpha\rangle\longrightarrow\frac{e^{-i\pi/4}}{\sqrt{2}}(|\alpha\rangle+i|-\alpha\rangle),\\
&&|-\alpha\rangle\longrightarrow\frac{e^{-i\pi/4}}{\sqrt{2}}(i|\alpha\rangle+|-\alpha\rangle),
\end{eqnarray}
which corresponds to $B_x$ up to a global phase shift.

The $B_z$ rotation can be obtained by displacement operator
$D(\delta)=\exp(\delta a^\dag-\delta^* a)$ \cite{JK}, where $a$ and
$a^\dag$ are respectively annihilation and creation operators.  We
know that two displacement operators $D(\alpha)$ and $D(\delta)$ do
not commute but the product $D(\alpha)D(\delta)$ is simply
$D(\alpha+\delta)$ multiplied by a phase factor, $\exp[{1\over
  2}(\alpha\delta^*-\alpha^*\delta)]$.  This phase factor plays a role
to rotate the logical qubit.  The action of displacement operator
$D(i\ep)$, where $\ep$ ($\ll 1$) is real, on a qubit
$|\phi\rangle=a|\alpha\rangle+b|-\alpha\rangle$ is the same as
$z$-rotation of the qubit by $U_z(\theta/2=2\alpha\ep)$.  We can
easily check their similarity by calculating the fidelity:
\begin{equation}
\label{fidelity-rotation}
|\langle \phi|U_z^\dag(2\alpha\ep)D(i\ep)|\phi\rangle|^2\simeq \exp[-\ep^2]\simeq1,
\end{equation}
where $\alpha\gg1$ is assumed and both $\alpha$ and $\ep$ are real.
Then the rotation angle can be represented as $\theta=4\alpha\ep$.  Note that a
small amount of $\ep$ suffices to make one cycle of rotation for a
large $\alpha$.
The displacement operation $D(i\ep)$ can be effectively performed
using a beam splitter with the transmission coefficient $T$ close to
unity and a high-intensity coherent field of amplitude $i{\cal E}$,
where ${\cal E}$ is real.  It is known that the effect of the beam
splitter is described by $D(i{\cal E}\sqrt{1-T})$ in the limit of
$T\rightarrow1$ and ${\cal E}\gg1$.  For the $B_z$ rotation, $\ep$ should
be taken to be $\pi/8\alpha$ so that the incident coherent field may
be $|i\pi/(8\alpha\sqrt{1-T})\rangle$.  The $B_y$ rotation can be realized by
applying $B_x$ and $B_z$ together with $\sigma_z$ noting
\begin{equation}
B_y=-\sigma_z B_x B_z B_x,
\end{equation}
where $\sigma_z$ is $\pi$ rotation around $z$ axis. 
The coherent state  $|i\pi/(4\alpha\sqrt{1-T})\rangle$ should be used
to perform $\sigma_z$.

Alice and Bob can perform random bilateral rotations
to transform the initial general mixed state into a
Werner state.  In this process, the efficiency of nonlinear
interaction can affect the efficiency of the scheme.

   \section{Purification for decohered states in vacuum} 
   \noindent

We now apply our scheme to a physical example in a dissipative environment. 
When the entangled coherent channel $|\Phi_-\rangle$ is embedded in a
vacuum, the channel decoheres and becomes a mixed state of
its density operator $\rho_{ab}(\tau)$, where $\tau$ stands for the
decoherence time.  By
solving the master equation \cite{Phoenix}
\begin{equation}
\label{master-eq}
{\partial \rho \over \partial \tau}=\hat{J}\rho +\hat{L}\rho~;~~
\hat{J}\rho=\gamma \sum_i a_i\rho a_i^\dag,~~
\hat{L}\rho=-{\gamma \over 2}\sum_i(a_i^\dag a_i\rho +\rho a_i^\dag a_i)
\end{equation}
where $\gamma$ is the energy decay rate, the mixed state
$\rho_{ab}(\tau)$ can be straightforwardly obtained as
\begin{eqnarray}
&&\rho_{ab}(\tau)={\wt N}(\tau)\Big\{|t\alpha, t\alpha \rangle\langle t\alpha, t\alpha|+|-t\alpha,
-t\alpha \rangle\langle -t\alpha, -t\alpha| \nonumber \\
&&~~~~~~~~~~~~~~~~~~~~~~-\Gamma(|t\alpha, t\alpha \rangle\langle -t\alpha, -t\alpha|+|-t\alpha, -t\alpha\rangle\langle t\alpha, t\alpha|
)\Big\},
\end{eqnarray}
where $|\pm t\alpha,\pm t\alpha\rangle=|\pm t\alpha\rangle_a|\pm
t\alpha\rangle_b$, $t=e^{-\gamma \tau/2}$,
$\Gamma=\exp[-4(1-t^2)|\alpha|^2]$, and ${\wt N}(\tau)$ is the normalization
factor.
The decohered state $\rho_{ab}(\tau)$ may be represented by the
dynamic quasi-Bell states defined as follows:
\begin{eqnarray}
\label{eq:dqbs1}
&&|{\wt \Phi}_\pm\rangle_{ab}={\wt N}_\pm(|t\alpha\rangle_a|t\alpha\rangle_b\pm|-t\alpha\rangle_a|-t\alpha\rangle_b),\\
\label{eq:dqbs2}
&&|{\wt \Psi}_\pm\rangle_{ab}={\wt
 N}_\pm(|t\alpha\rangle_a|-t\alpha\rangle_b\pm|-t\alpha\rangle_a|t\alpha\rangle_b),
\end{eqnarray}
where ${\wt N}_\pm=\{2(1\pm e^{-4t^2|\alpha|^2})\}^{-1/2}$.
The decohered state is then
\begin{eqnarray}
\label{decs}
&&\rho_{ab}(\tau)={\wt N}(\tau)\big\{\frac{(1+\Gamma)}{{\wt
  N}_-^2}|{\wt \Phi}_-\rangle\langle{\wt \Phi}_-|+\frac{(1-\Gamma)}{{\wt
  N}_-^2}|{\wt \Phi}_+\rangle\langle{\wt \Phi}_+|\big\} \nonumber   \\
\label{decay}
&&~~~~~~~~\equiv F(\tau)|{\wt \Phi}_-\rangle\langle{\wt \Phi}_-|+(1-F(\tau))|{\wt \Phi}_+\rangle\langle{\wt \Phi}_+|
\end{eqnarray}
where, regardless of the decay time $\tau$, $|{\wt \Phi}_-\rangle$ is 
maximally entangled and $|{\wt \Phi}_-\rangle$ and  $|{\wt \Phi}_+\rangle$
are orthogonal to each other.
 The decohered state (\ref{decay}) 
is not in the same form as Eq.~(\ref{ensemble}) so that we need some
bilateral unitary transformations before the purification scheme is applied. 
We find that unitary operation $B_x$ on each side of Alice and Bob
transforms the state into
\begin{equation}
\label{ts}
B_{xa} B_{xb}\rho_{ab}(\tau)B_{xb}^\dagger B_{xa}^\dagger=F(\tau)|{\wt \Phi}_-\rangle\langle{\wt \Phi}_-|+(1-F(\tau))|{\wt \Psi}_+\rangle\langle{\wt \Psi}_+|,
\end{equation}
which is obviously  
distillable form using the schemes explained in Sec.~3.
A Hadamard gate $H$ for coherent states \cite{JK,Ralph} can also be used to transform the state (\ref{decs}) into a distillable form
\begin{equation}
\label{ts2}
H_a H_b\rho_{ab}(\tau)H_b^\dagger H_a^\dagger \nonumber \\
=N_h\Big\{F(\tau)|{\wt \Psi}_+\rangle\langle{\wt \Psi}_+|+(1-F(\tau))|{\wt \Phi}_+\rangle\langle{\wt \Phi}_+|\Big\},
\end{equation}
where $N_h$ is the normalization factor due to the nonzero overlap
between $|{\wt \Psi}_+\rangle$ and $|{\wt \Phi}_+\rangle$.  Note that
the Hadamard operation for coherent states can be approximately
realized using linear elements if cat states
are pre-arranged \cite{Ralph}.

The ensemble of state (\ref{decay}) can be purified successfully only when $F(\tau)$ is
larger than 1/2. Because
\begin{equation}
F(\tau)=\frac{N_+^2(1+\Gamma)}{N_+^2(1+\Gamma)-N_-^2(1-\Gamma)},
\end{equation}
it is found that purification is successful
when the decoherence time $\gamma\tau<\ln 2$ regardless of $\alpha$.  This result is in agreement with
the decay time until which teleportation can be successfully
performed via an entangled coherent state shown in ref. \cite{JKL01}.

   \section{Multi-mode purification} 
   \noindent 
 
   Besides a two-mode entangled coherent state, a multi-mode entangled
   coherent state \cite{WS} can be used for quantum computation using
   coherent-state qubits \cite{JK}.  There is a suggestion for
   multi-mode entanglement purification based on controlled-NOT
   operation \cite{Murao}.  In this section we investigate an example
   of application of our scheme to multi-mode entangled states.

 \begin{figure} [htbp] 
   \vspace*{13pt} 
   \centerline{\psfig{file=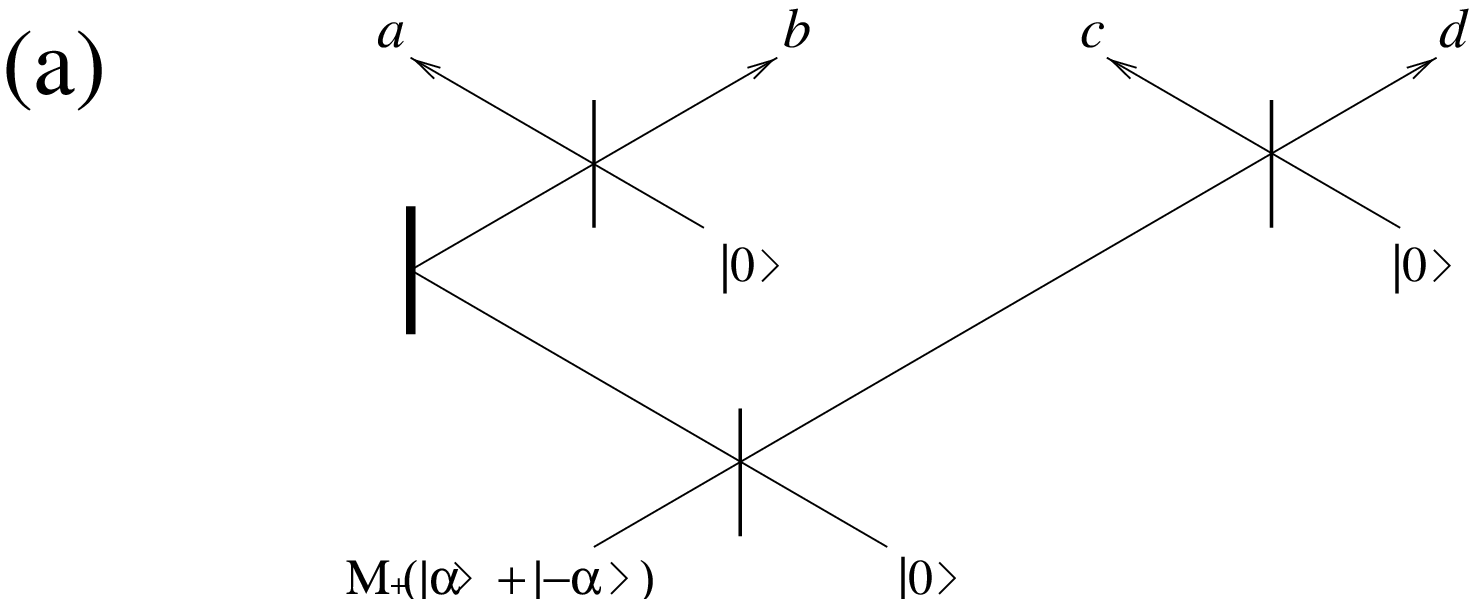, width=6.4cm}} 
   \vspace*{13pt} 
   \centerline{\psfig{file=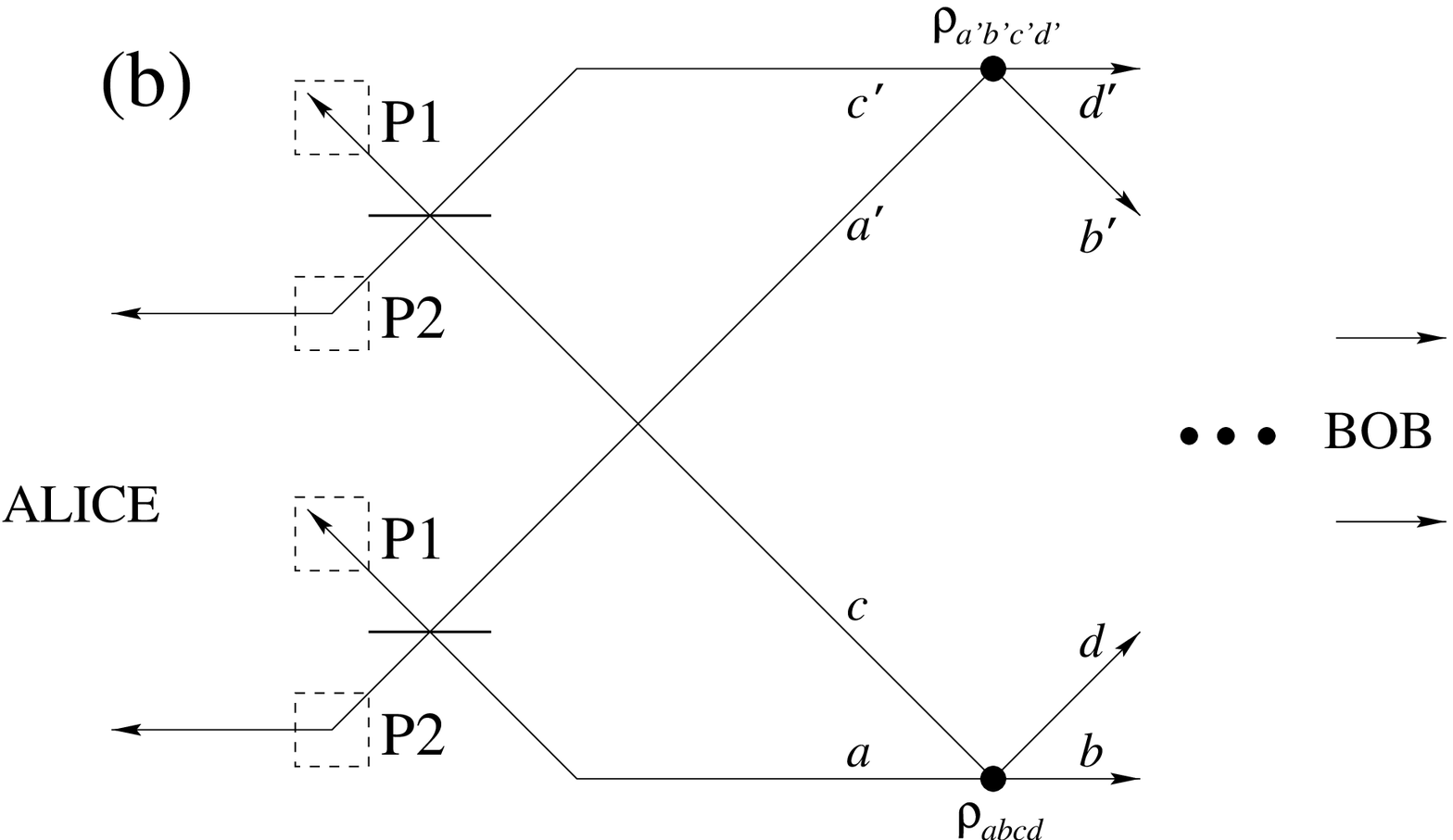, width=7.5cm}} 
   \vspace*{13pt} 
   \fcaption{
(a) Schematic for generation of a four-mode entangled coherent state using
  a nonlinear medium and 50-50 beam splitters.  
  A coherent-superposition state $M_+(|\alpha\rangle+|-\alpha\rangle)$
  can be prepared using a nonlinear medium before it passes through
  beam splitters.  (b) Entanglement purification for four-mode
  entangled coherent
  states.
\label{multi} } 
   \end{figure}

Multi-mode entangled coherent states can be generated using a
coherent superposition state and 50-50 beam splitters. The number of
required beam splitters is $N-1$, where $N$ is the number of modes for the
multi-mode entangled state.  For example, a four-mode entangled state
can be generated as shown in Fig.~\ref{multi}(a). After passing the
three beam splitters, the four-mode entangled state
$|B_1\rangle={\cal
  N}_-(|\alpha,\alpha,\alpha,\alpha\rangle+|-\alpha,-\alpha,-\alpha,-\alpha\rangle)$ is generated.
Suppose Alice and Bob's ensemble to be purified is represented by
\begin{equation}
\label{eq:rho4m}
\rho_{ab}=F|B_1\rangle\langle B_1|+G|B_2\rangle\langle B_2|,
\end{equation}
where $|B_2\rangle={\cal
  N}_-(|\alpha,-\alpha,\alpha,-\alpha\rangle+|-\alpha,\alpha,-\alpha,\alpha\rangle)$ and $|B_2\rangle$ can be generated in a similar way as
$|B_1\rangle$. By extending the scheme studied above, the ensemble (\ref{eq:rho4m}) can be
purified as shown in Fig.~\ref{multi}(b).
After one successful iteration of the purification process, the originally
  selected pairs become 
\begin{equation}
\label{eq:rho4mi}
\rho_{ab}=F^\prime|B_1\rangle\langle B_1|+G^\prime|B_2\rangle\langle B_2|,
\end{equation}
where $F^\prime=\frac{F^2}{F^2+(1-F^2)}$ is always larger than $F$ for
$F>G$.  Alice and Bob can iterate the process as many time as required
for their use.  Note that this scheme can be applied to any $N$-mode
entangled states of the same type and so does the simpler scheme only with P1.

   \section{Remarks} 
   \noindent 

   We have suggested an entanglement purification scheme for mixed
   entangled coherent states.  Our scheme is based on the use of 50-50
   beam splitters and photodetectors.  The scheme is directly
   applicable for mixed entangled coherent states of the Werner type,
   and can be useful for general two-mode mixed states using
   additional nonlinear interactions.  We have also suggested a
   simplified variation of this scheme which, however, increases the
   coherent amplitude of the entangled coherent state.  We applied our
   scheme to an entangled coherent state decohered in a vacuum
   environment.

   Finally, we would like to address possible difficulties for
   experimental realization of the purification scheme.  
   We already pointed out that the nonlinear interaction required for
   the generation of cat states and for additional bilateral rotations
   to purify some non-Werner type states is extremely difficult, while
   the efforts 
to improve nonlinearity that additional noise is being continuously
   investigated. 
  If the entangled coherent state is subject to a
   thermal environment, it is not straightforward to represent the decohered
   state in
   the simple basis of the quasi-Bell states 
   (\ref{eq:qbs1}) and (\ref{eq:qbs2}).  
The purification of a decohered state due to thermal environment will
be much more complicated and will deserve further studies.
Laser phase drift can be another obstacle to
realize quantum information processing with coherent states 
and phase stabilization methods via mixing of laser beams can be used
to reduce the drifts.  Apart from the phase drift, 
There has been a controversy on 
whether conventional laser sources can be used for quantum
communication 
with coherent states \cite{Rudolph,EF}.  Most
recent study \cite{EF} shows that 
the conventional laser can be used for quantum teleportation and for
generating continuous-variable entanglement because optical coherence
is not necessary for the purpose.

For quantum information processing, an entangled coherent state is
normally assumed to have a large coherent amplitude.  Even though the
success probability of the purification scheme is better as the
coherent amplitude, $\alpha$, is larger, it does not change much.  For
example, the success rate is about 5\% degraded for $\alpha=1$ compared
with the case for $\alpha\rightarrow\infty$.  To use the first
purification scheme described in Fig. 2(a), even and odd numbers of
photons have to be discriminated. If the coherent amplitude is large,
the efficiency to discriminate even and odd numbers of photons becomes
low due to the dark current.  However, when the coherent amplitude is
small, a highly efficient avalanche photodiode can be used to
discriminate even (0 and 2) photon numbers and odd (1) photon number
\cite{Takeuchi} because, for example, taking $\alpha= 1$ the probability of
photon number being 0 and 2 for an even cat state is about 97\% and
the probability of photon number being 1 is about 85\%.  Takeuchi {\em
  et al}. used threefold tight shielding and viewports that worked as
infrared blocking filters to eleminate the dark count.  On the other
hand, the second purification scheme in Fig. 2(b) is robust against
detection inefficiency when $\alpha$ is large because it is enough to
discern a coherent state and a vacuum in this simplified scheme.  By
employing a distributed photon counter or a homodyne detector, we have
even a higher detection efficiency to discern a coherent state and a
vacuum.

\nonumsection{Acknowledgements}
   \noindent 

We acknowledge stimulating discussions with 
J. Lee, and helpful advice of E. Solano, N. Korolkova and N. Imoto.
We thank the UK Engineering and Physical Sciences Research Council for
financial support through GR/R33304. HJ acknowledges the Overseas
Research Student award.

   \nonumsection{References}

   \end{document}